\documentclass[12pt]{article}
\usepackage{amssymb,latexsym}
\title{Entropy of Isolated Horizons revisited}
\author {Rudranil Basu \\ {\it SN Bose National
Centre for Basic Sciences, Kolkata 700 098, India} \\ Romesh K. Kaul \\ {\it The Institute of Mathematical
Sciences, Chennai 600 113, India}  \\
Parthasarathi Majumdar \\ {\it Saha Institute of Nuclear
Physics, Kolkata 700 064, India }}
\begin{document}
\maketitle
\begin{abstract}

The decade-old formulation of the isolated horizon classically and within loop
quantum gravity, and the extraction of the microcanonical entropy of such a
horizon from this formulation, is reviewed, in view of recent renewed
interest.  There are two main approaches to this problem: one employs 
an $SU(2)$ Chern-Simons theory describing the isolated horizon
degrees of freedom, while the other uses a reduced $U(1)$ Chern-Simons theory
obtained from the $SU(2)$ theory, with appropriate constraints imposed on the
spectrum of boundary states `living' on the horizon. It is shown that both
these ways lead to the same infinite series asymptotic in horizon area for
the microcanonical entropy of an isolated horizon. The leading area term is
followed by an unambiguous correction term
logarithmic in area with a coefficient $-\frac32$, with subleading corrections
dropping off as inverse powers of the area.   
\end{abstract}

\section{Introduction}

There appears to have been a resurgence in interest in the Loop Quantum Gravity
approach towards black hole entropy. The main idea of this approach involves
identifying a `boundary' theory characterizing the degrees of freedom on an
isolated horizon (of fixed cross-sectional area), consistent with the boundary conditions used to
{\it define} such horizons \cite{abf} \cite{aba}, and then counting the dimension of the Hilbert
space of the quantum version of this boundary theory \cite{abck, kmplb, kmprl,
abck2}. This
dimension is then considered to be the exponential of the {\it microcanonical}
entropy of the isolated horizon. Clearly this is an effective field theory approach where 
the existence of an isolated horizon, as a null inner boundary of quantum
space (on a spatial slice) punctured by spin
network links in loop quantum gravity, is {\it assumed} from the
start, and not derived as a solution of the quantum Einstein equation (the
Hamiltonian constraint, in a canonical description). Thus, one has to further
make the assumption that the quantum Einstein equation does indeed yield
spacetimes with this assumed property.

Before we begin, a word about our {\it notation} : spacetime forms which are also internal $SU(2)$ vectors are
indicated with a boldface and an arrow on top. The exterior product between
such forms includes the cross product between the internal vectors. The
exterior product with an overdot indicates a standard exterior product between the
spacetime forms and a scalar product between the internal $SU(2)$ vectors. For
spacetime forms which are not internal vectors, indicated in boldface without
arrow on top, the exterior product has the standard connotation. A plain
cross-product operates between internal vector functions or between a function
and a form which is also an internal vector.

Within a canonical formulation, vacuum general relativity is formulated on a
partial Cauchy surface $M$ in
terms of the Barbero-Immirzi (BI) class of $SU(2)$ Lie-algebra valued connection
one-forms expressed in the basis of $SU(2)$ generators ${\vec {\bf A}}$.\footnote{The reduction in gauge
invariance from the
full local Lorentz group ($SL(2,C)$) to the group of local rotations ($SU(2)$) is made by fixing the
so-called time gauge whereby local Lorentz boosts are fixed on the spatial
slice.} The canonically conjugate phase space variable is the $SU(2)$-valued solder
form, expressed in the basis of $SU(2)$ generators ${\vec {\bf
\Sigma}}$. In terms of these, the symplectic structure of
vacuum spacetime takes the form (ignoring boundary terms) 
\begin{eqnarray}
\Omega_V(\delta_1,\delta_2) = {1 \over 16\pi G} \int_M \delta_{[1} {\vec {\bf A}}
{\dot \wedge} \delta_{2]} {\vec {\bf \Sigma}}~. \label{symv}
\end{eqnarray} 
  
We now introduce an isolated horizon as a null {\it inner} boundary of
spacetime with fixed cross sectional area $A_{IH}$ \cite{abf}. The boundary conditions
do not lead to any obvious reduction in gauge invariance, so one expects the
boundary theory to be a three dimensional topological gauge theory living on
the null surface and having an $SU(2)$ gauge invariance. This theory is 
expected to be an $SU(2)$ Chern-Simons theory. An equivalent alternative
description of the boundary theory in terms of a topological $SU(2) ~{\bf B}-{\bf F}$
theory is also possible. We should perhaps mention that the use of
$SU(2)$ Chern-Simons theory as a boundary theory to derive black hole entropy
has a precedence: independently, Krasnov, Rovelli and Smolin \cite{krs}
have considered this possibility, although not using the Isolated
Horizon paradigm. The task of
implementing the boundary conditions and deriving the phase space symplectic
structure is nontrivial and has been accomplished with several simplifying
assumptions, involving an additional gauge fixing \cite{abck}, which reduces
the gauge invariance to $U(1)$. As for any fixing of gauge,
the convenience is always accompanied by additional constraints on the
dynamical variables like the curvature; unfortunately, these constraints are
not always implemented fully when
deriving the symplectic structure. If they are taken into account, one expects
to regain the full $SU(2)$ gauge invariance in phase space. Work is in
progress to demonstrate this explicitly from generic isolated horizon boundary
conditions \cite{bm} (see, however, the recent preprint \cite{per} which
discusses an $SU(2)$ CS theory as the theory on an IH)). 

In this short note, we first {\it assume} that the topological gauge theory on the
isolated horizon is indeed an $SU(2)$ Chern-Simons theory with a coupling
constant $k \equiv A_{IH} / 8\pi \gamma l_P^2$. To be consistent with known
properties of $SU(2)$ Chern-Simons theory, one assumes that $A_{IH} >> l_P^2$
and the nearest integer value of $k$ to the expression above is chosen. We
then briefly review the
derivation, given in 1998 by two of us, of the dimensionality of the
Hilbert space of the quantum version of the theory \cite{kmplb}; in that paper
{\it $SU(2)$ singlet states}  are counted 
using the relation \cite{wit} between the Hilbert space of an
$SU(2)$ Chern-Simons theory on a punctured $S^2 \times {\bf R}$ with the
number of conformal blocks of the $SU(2)_k$ WZW model `living ' in that $S^2$.  
We also review the further derivation by two of us \cite{kmprl} (given in 2000)
showing how this leads to an infinite series asymptotic in
$A_{IH}$ for the microcanonical entropy, with a leading area term and
corrections beginning with a term logarithmic in the area with a
coefficient $-3/2$. 

Next we do the
counting of states differently : by first gauge fixing the $SU(2)$ to an $U(1)$ following the
pioneering work of Ashtekar {\it et. al.} \cite{abck} using a covariantly constant
internal vector ${\vec r}$. One then counts the states that are $U(1)$
invariant. However, we consider
additionally the constraint on the $SU(2)$ curvature that the gauge fixing
entails. At the level of the quantum theory, these additional constraints are
shown to lead to precisely the same counting of states as recalled above in
terms of $SU(2)$ invariant states. There is thus no discrepancy in the
final answer for the microcanonical entropy. 

\section{Review of $SU(2)$ singlet state counting}

In presence of the isolated horizon the bulk symplectic structure above is
augmented by a boundary symplectic structure assumed to be given by that of an
$SU(2)$ Chern-Simons theory,
\begin{eqnarray}
\Omega_b(\delta_1,\delta_2) = {k \over 2\pi}  \int_S \delta_{[1} {\vec {\bf A}}
{\dot \wedge} \delta_{2]} {\vec {\bf A}} 
\end{eqnarray}
where $k \equiv A_{IH} /2\pi \gamma$ with $\gamma$ being the Barbero-Immirzi
parameter. Here $S$ is the spatial $S^2$ foliation of the isolated horizon. On
this $S^2$ the `Gauss law' equation appropriate to local spatial rotations is
\begin{eqnarray}
{k \over 2\pi} {\vec {\bf F}} ~=~- {\vec {\bf \Sigma}} ~. \label{key}
\end{eqnarray}
In eq. (\ref{key}), both ${\vec {\bf F}}$ and ${\vec {\bf \Sigma}}$ are pull backs of the
curvature 2-form appropriate to the Barbero-Immirzi connection and the solder form on
partial Cauchy surface $M$ to $S$. 

In LQG, the Hilbert space is assumed to be the tensor product ${\cal H}_V
\otimes {\cal H}_{IH}$ corresponding to the bulk spin network space and the
isolated horizon respectively. The geometric variables in (\ref{key}) above
become operators acting on appropriate Hilbert spaces. The solder form
operator has an action on spin network bulk states as eigenstates with
support only on the punctures of $S$ where spin network links pierce
it. Consequently, acting on the isolated network boundary states, the
curvature operator has the action
\begin{eqnarray}
{\cal I} \otimes {k \over 2\pi} {\hat {\vec {\bf F}}}(x) |\psi_V\rangle \otimes | \chi_{IH}\rangle = - \sum_p
\delta^{(2)}(x,x_p) ~{}^2 \epsilon_p ~{\vec T}_p |\psi_V \rangle \otimes |\chi_{IH} \rangle    
\end{eqnarray}
where, the sum is over a set of punctures carrying $SU(2)$ spin representation
of the generators ${\vec T}_p$ (corresponding to spin $j_p$) at the $p$th
puncture, and ${}^2\epsilon_p$ is the area 2-form for that
puncture. Given that upto $O(l_P^2)$, the sum of these areas over the entire
set of punctures must equal the fixed classical area $A_{IH}$, one immediately
realizes that the set of states to be counted must obey the constraint that
they are {\it $SU(2)$ singlets}. 

This counting has been accomplished in \cite{kmplb}. One utilizes the
connection \cite{wit} between the dimensionality of the Hilbert space of $SU(2)$
Chern-Simons theory living on
a punctured $S^2$ ($\times {\cal R}$) and the number of {\it conformal
  blocks} of the boundary two dimensional conformal field theory --  $SU(2)_k$
WZW model on the punctured $S^2$. One also makes use of the fusion algebra and
the Verlinde formula for the representation matrices of that algebra.   
In terms of the spins $j_1, j_2, \dots, j_p$
on punctures, the dimension of the space of $SU(2)$-singlet boundary states is
(for $k \rightarrow \infty$)  
\begin{eqnarray}
{\cal N}(j_1, \dots , j_p) =  \prod_{n=1}^p \sum_{m_n=-j_n} ^{j_n} \left
  (\delta_{m_1 + \cdots+ m_p, 0} 
- \frac12 \delta_{m_1 + \cdots + m_p, -1} - \frac12 \delta_{m_1 + \cdots + m_p, 1} \right)
~. \label{dimen}   
\end{eqnarray}
The last two terms precisely ensure that the counting is
restricted to $SU(2)$ singlet boundary states, since these alone obey the
`Gauss law constraint' which ensures local gauge invariance or `physicality'
of the counted states.

To extract the microcanonical entropy of the isolated horizon, one may follow
our work \cite{kmprl}; the entropy turns out to be 
\begin{equation}
S_{IH}~=~S_{BH}~-~{3 \over 2}~\log S_{BH}~+~const.~+~O(S_{BH}^{-1}), \label{main}
\end{equation}
where, $S_{BH}$ is the usual Bekenstein-Hawking area expression for the entropy:
$S_{BH} = A_{IH} /4 l_P^2$. In
this, the Barbero-Immirzi parameter has been `fitted' to agree with the
correct normalization of the Bekenstein-Hawking area term. There is absolutely
no other ambiguity in this infinite series, each of whose terms are finite and
calculable.  

\section{The $U(1)$ counting}

The implementation of the isolated horizon boundary conditions and derivation
of the boundary symplectic structure has been accomplished at the classical
level in \cite{abf} and later follow-up work \cite{aba}. It has been claimed that this is
most facile if one makes a further fixing of the gauge invariance on a Cauchy
surface from $SU(2)$ to a residual $U(1)$ invariance generated by the diagonal
$SU(2)$ generator alone. This is done by picking an internal $SU(2)$
vector field ${\vec r}$ which is covariantly constant on the $S^2$ foliation
of an isolated horizon (or equivalently an $su(2)$-valued
function which is covariantly constant on the $S^2$). It is obvious such an
internal vector field always exists on the $S^2$.  
\begin{eqnarray}  
D {\vec r} \equiv d{\vec r} + {\vec {\bf A}} \times {\vec r} = 0. \label{covr}
\end{eqnarray}
where ${\vec {\bf A}}$ is pull-back to the $S^2$ of the $SU(2)$ BI connection

\begin{eqnarray}
{\vec {\bf A}} = {\vec r}{\bf B} + {\vec {\bf C}} \label{dec}
\end{eqnarray}
with,
\begin{eqnarray}
{\vec r} \cdot {\vec {\bf C}} &=& 0~,~{\vec r}^2 =1 \nonumber \\
D {\vec r} &= & d{\vec r} + {\vec {\bf C}} \times {\vec r} = 0~. \label{con} 
\end{eqnarray} 
Observe that one can solve the second equation above explicitly for ${\vec {\bf C}}$
\begin{eqnarray}
{\vec {\bf C}} &=& - {\vec r} \times d{\vec r} ~. \label{cee} 
\end{eqnarray}
The pullback of the curvature two-form to the $S^2$ is 
\begin{eqnarray}
{\vec {\bf F}} & = & d{\vec {\bf A}} + \frac12 {\vec {\bf A}} \wedge {\vec {\bf A}}
\nonumber \\
&=& {\vec r} \left( d{\bf B} - \frac12 {\vec r} \cdot d{\vec r} \wedge d{\vec
    r} \right) 
\label{efff}
\end{eqnarray}
The second line of eq. (\ref{efff}) follows from the first using the decomposition
(\ref{dec}) and the solution (\ref{cee}), together with the observation that
the 2-form $d{\vec r} \wedge d{\vec r} = {\vec r}~{\vec r} \cdot d{\vec
  r} \wedge d{\vec r}$ which ensues because ${\vec r} \times d{\vec r} \wedge
d{\vec r} =0$.   

The projection of this curvature along ${\vec r}$ is given by
\begin{eqnarray}
{\bf f} \equiv {\vec r}\cdot {\vec {\bf F}} = d{\bf B} - \frac12 {\vec r} \cdot d{\vec r}
\wedge d{\vec r} ~. \label{r.f}
\end{eqnarray}
The second term in eq. (\ref{r.f}) is actually a winding number density
associated with maps from $S^2$ to $S^2$; if we write it as $- d{\bf \Omega}$, then 
\begin{eqnarray}
{1 \over 8\pi} \int_S d{\bf \Omega} = N \in {\cal Z} ~. 
\end{eqnarray}
Thus, we may write the $U(1)$ curvature as
\begin{eqnarray}
{\bf f} = d{\bf B'} \label{r.f2}
\end{eqnarray}
where, ${\bf B'} \equiv {\bf B} - {\bf \Omega}$. We note that for the quantum
isolated horizon, the $U(1)$ connection
${\bf B'}$ vanishes locally on the $S^2$, except on the punctures.
Because of the nontrivial winding at each
puncture, it is a nontrivial $U(1)$ bundle on $S^2$. This is the contribution
that accumulates to giving the counting of states leading to the
microcanonical entropy in this approach. 

The counting now proceeds by solving the $U(1)$ projected version of
eq. (\ref{key})
\begin{eqnarray}
{k \over 2\pi} {\bf f} = - {\vec r} \cdot {\vec {\bf \Sigma}} ~\label{prkey} 
\end{eqnarray}
which immediately translates, in the quantum version of the theory to 
\begin{eqnarray}
{\cal I} \otimes {k \over 2\pi} {\hat {\bf f}}(x) |\psi_V\rangle \otimes | \chi_{IH}\rangle_{U(1)} = - \sum_p
\delta^{(2)}(x,x_p) {}^2 \epsilon_p ~ {\vec r} \cdot {\vec T}_p |\psi_V \rangle
\otimes |\chi_{IH} \rangle_{U(1)}    ~, \label{u1cou}
\end{eqnarray}
which implies that the states to be counted are $U(1)$ Chern-Simons theory
states on the punctured sphere with the net spin projection along ${\vec r}$
vanishing: ${\vec r} \cdot \sum_p {\vec T}_p=0$. Observe that one can always
rotate ${\vec r}$ locally so that this is possible, even though globally this
vector corresponds to a nontrivial $U(1)$ bundle on $S$. 

This $U(1)$ counting has been done in a variety of ways \cite{u1md}.
If, in the reduced $U(1)$ theory, we disregard the consequences
of the constraint implied for the projection of $SU(2)$ field
strength orthogonal to the direction of vector $\vec r$ (see the next paragraph),
the final result for the dimensionality of the $U(1)$ Chern-Simons Hilbert
space is given by the first term of the eqn. (\ref{dimen}).
For macroscopic ($A_{IH} >> l_P^2$) isolated
horizons the corresponding microcanonical entropy is given by
\begin{eqnarray}
S_{IH} = S_{BH} -\frac12 \log S_{BH} + \cdots ~. \label{u1ent}
\end{eqnarray}
The leading term once again offers a fit to the BI parameter, which is the
same as in the $SU(2)$ case, if spins at all punctures are chosen to be $1/2$.
The most obvious difference though is
the appearance of a logarithmic LQG correction to the Bekenstein-Hawking area
term, {\it with a coefficient $-1/2$} instead of $-3/2$ as found above by doing the
$SU(2)$ singlet counting. This is obviously because of the additional gauge
fixing performed in implementing the isolated horizon boundary conditions; the
diagonal $SU(2)$ generator is taken parallel to the covariantly constant
internal vector field
${\vec r}$ chosen above. Thus, the generators orthogonal to ${\vec r}$ are
set to zero, and hence the apparent discrepancy between (\ref{u1ent}) and (\ref{main}). 

However, observe that the curvature given in (\ref{efff}) has {\it
  vanishing} projection orthogonal to ${\vec r}$, i.e., 
\begin{eqnarray}
{\vec {\bf F}} \times {\vec r} = 0~. \label{orth}
\end{eqnarray}
This can also be derived independently because of the gauge choice in terms
of the special internal vector
${\vec r}$ obeying $D({\vec {\bf A}}) {\vec r} =0$; one obtains the
same constraint
\begin{eqnarray}
[D_a, D_b] {\vec r} = 0 = {\vec F}_{ab} \times {\vec r} 
\end{eqnarray}
where $a,b$ are spacetime indices on $S^2$. This constraint arises as an
essential and {\it inevitable} part of the
additional gauge fixing performed on the theory on $S$, reducing the residual
invariance on $S$ (in time gauge) from $SU(2)$ to $U(1)$. The constraint imposes a
direct and very significant additional restriction on the class of `physical'
states contributing to the microcanonical entropy, over and above
that of $U(1)$-neutrality. If we use eq. (\ref{key}) and consider the quantum
version of the above additional constraint on the spin network bulk and
boundary states, we obtain,
\begin{eqnarray}
\sum_p \delta^{(2)}(x,x_p)~{}^2\epsilon_p~{\vec r} \times {\vec T_p}
|\psi_V\rangle \otimes |\chi_{IH} \rangle_{U(1)} = 0 \label{add}
\end{eqnarray}    
where ${\vec T}$ are the $su(2)$
generators. One must now count the dimension of IH states that satisfy the
additional constraint (\ref{add}) apart from $U(1)$ neutrality. {\it This,
  unfortunately, has not been done in the later literature on $U(1)$ counting
  approaches \cite{u1md}}. 

Consider now the behaviour of eq. (\ref{u1cou}) under the action of the $SU(2)$
Gauss law constraint. Under the action of this
constraint the states are annihilated. It is easy to check that the LHS of
(\ref{u1cou}) is actually invariant under $SU(2)$ gauge transformations (for a
fixed ${\vec r}$, as in this case)
generated by the Gauss law constraint, provided one makes use of the operator
version of (\ref{orth}). More specifically, under an $SU(2)$ gauge
transformation, $ {\vec {\bf F}} \rightarrow {\vec {\bf F}}' = {\vec {\bf F}}
+ {\vec \theta} \times {\vec {\bf F}}$ where ${\vec \theta}$ is an
infinitesimal transformation parameter. Note though that the U(1) field
strength ${\bf f}$ does not change, because of (\ref{orth})
\begin{eqnarray}
{\bf f}' &=&  {\vec r} \cdot {\vec {\bf F}} + {\vec r} \cdot {\vec \theta}
\times {\vec {\bf F}} \nonumber \\
&=& {\bf f} + {\vec {\theta}} \cdot {\vec {\bf F}} \times {\vec r} = {\bf f}
\end{eqnarray}
The LHS of (\ref{u1cou}) thus remains invariant under local $SU(2)$ gauge
transformations. There is a small subtlety in the manner in which the RHS of
(\ref{u1cou}) also remains invariant under such. Observe that such transformations must
not remove the system off
the gauge surface defined by the gauge fixing condition (\ref{con}), i.e.,
$D{\vec r}=0$, valid on the 2-sphere foliation of the IH. This imposes the
restriction on the $SU(2)$ gauge parameter ${\vec \theta}$ to lie along the
fixed internal vector ${\vec r}$, i.e., ${\vec \theta} = {\vec r} \alpha$ on
such 2-sphere foliations. The parameter can of course be arbitrary away from
the 2-spheres. It is easy to see that the RHS of (\ref{u1cou}) is indeed
invariant under such restricted $SU(2)$ transformations, as also is the
additional constraint eq. (\ref{add}) on the IH states, both being relations
valid on the 2-sphere foliations $S$.      

To conclude, note that the import of eqn. (\ref{add}) is to imply that spins on 
punctures should be such that the net spin orthogonal to the
direction $\vec r$ is zero. This then, along with net $U(1)$ neutrality,
is the requirement that all admissible isolated horizon states
contributing to the microcanonical entropy are $SU(2)$ singlets.
This leads to the LQG result for the entropy as written out in the formula
(\ref{dimen}), where the first term on the right-hand side counts the
$U(1)$ neutral states and the last two terms subtract out the
overcounted states with the net azimuthal quantum number equal to zero coming from
net non-zero
spin ($1, ~2, ~3, ......$) states which need to be excluded to impliment the
additional constraint (\ref{add}). Thus recalling (6), even in this counting
the leading logarithmic LQG correction
has a coefficient $ - {\frac 32}$ as  advertized in the abstract and
in the introduction.

\vglue 1cm

\noindent {\bf Acknowledgement :} One of us (RKK) thanks G. Date for helpful
discussions. He also gratefully acknowledges the hospitality at the Centre for
High Energy Physics, Indian Institute of Science, Bangalore where this work
was completed.

\end{document}